\documentstyle[psfig,epsfig]{lamuphys}
\makeatletter

\let\chapter\hid@chapter
\makeatother

\begin{document}
\pagenumbering{arabic}
\title{Transport in semiconductor superlattices:
from quantum kinetics to terahertz-photon detectors}

\author{A. P. Jauho$^1$, A. Wacker$^2$, and A. A. Ignatov$^{1,3}$}

\institute{
$^1$Mikroelektronik Centret,
Technical University of Denmark, Bldg. 345east,
DK-2800 Lyngby, Denmark\\
$^2$Institute f{\"u}r Theoretische Physik, Technische Universit{\"a}t
Berlin, Hardenbergstr. 36, 10623 Berlin, Germany\\
$^3$Institute for Physics of Microstructures, Russian Academy of Science,
603600 Nizhny Novgorod, Russia}
\titlerunning{Transport in semiconductor superlattices}
\maketitle

\begin{abstract}

Semiconductor superlattices are interesting for two distinct
reasons:  the possibility to design their structure (band-width(s),
doping, etc.) gives access to a large parameter space where different
physical phenomena can be explored. Secondly, many  
important device applications have been proposed, and
then subsequently successfully fabricated.  A number of
theoretical approaches has been used to describe their
current-voltage characteristics, such as miniband conduction,
Wannier-Stark hopping, and sequential tunneling.  The choice
of a transport model has often been dictated by pragmatic
considerations without paying much attention to the strict domains
of validity of the chosen model.  In the first part of this
paper we review recent efforts to map out these
boundaries, using a first-principles quantum transport theory,
which encompasses the standard models as special cases.   In the
second part, focusing in the mini-band regime,
we analyze a superlattice device as an element
in an electric circuit, and show that its performance as
a THz-photon detector allows significant optimization, with
respect to geometric and parasitic effects, and detection
frequency.  The key physical mechanism enhancing the responsivity
is the excitation of hybrid Bloch-plasma oscillations.
 
\end{abstract}
\section{Introduction}

Ever since the pioneering work of Esaki and Tsu \cite{ESA70},
which drew attention to the rich physics and
potential device applications of semiconductor
superlattices, these man-made structures have remained
a topic of intense research.
Semiconductor superlattices have proven to be a
fruitful platform for studying a wide range of transport
phenomena, such as their intrinsic negative differential 
conductivity \cite{SIB90},  
the formation of electric field domains \cite{GRA91}, 
Bloch oscillations \cite{WAS93}, as well as dynamical 
localization \cite{HOL92} and
absolute negative conductance \cite{KEA95b} under external irradiation, 
just to mention a few.

These phenomena depend crucially on the relations of the
energy scales involved,
namely the zero-field miniband width (which is four times 
the interwell coupling $T_1$), the scattering rate $\Gamma/\hbar$,
and the potential drop per period ($\equiv eFd$, where $F$ is the 
applied static field and $d$ is the superlattice period).
Three distinct approaches
have been used to describe transport in the parameter
space spanned by $(T_1,eFd,\Gamma)$: 
miniband conduction (MBC)\cite{ESA70,LEB70},
Wannier-Stark hopping (WSH)\cite{TSU75}, and sequential 
tunneling (ST)\cite{MIL94,WAC97b}.
Until recently, however, the precise range of validity of these
various approaches had not been addressed quantitatively.
To achieve this, one much use a first-principles approach, which
reduces to the standard theories in the appropriate limits.
After a brief review of the standard approaches in Sect. II,
we introduce in Sect. III a nonequilibrium Green function formalism, which
we have used to delineate the boundaries of the different domains
of validity \cite{prl1,prl2}.  
We also present  comparisons with Monte Carlo
simulations.   As we shall see, under favorable conditions it is
quite possible to obtain a very accurate description with the standard
methods, which is a great advantage because the first-principle
Green function calculations are quite complicated and have so far
successfully implemented only for rather simple scattering
mechanisms (the scattering matrix elements are assumed to be
independent of momentum transfer).  Thus, in Sect. IV we adopt one of the standard approaches,
i.e., the miniband approach, to model a superlattice THz-photon detector,
taking into account the effects due to the external circuitry.  We conclude
that by detailed modeling substantial device performance optimization
can be achieved, e.g. the detector sensivity may be improved by almost
50 \% by a judicious choice of its parameters.

\section{Standard transport models}

Here we review the standard models used to describe transport in 
semiconductor superlattices.  For simplicity, the results quoted
in the next three subsections
are written in a relaxation time approximation, but a generalization to
more realistic scattering processes is possible.  As an underlying
Hamiltonian we use a tight-binding model:
\begin{equation}
\hat{H}^{\rm SL}_{n,m}=\left(\delta_{n,m-1}+\delta_{n,m+1}\right)T_1
+\delta_{n,m}(E_k-neFd)\label{Eqhamz} \, .
\end{equation}
Here $T_1$ is the overlap matrix element, $E_k=\hbar^2 k^2/(2m)$ is
the kinetic energy perpendicular to the growth direction, and the electric field
is taken into account by a shift in the site energies.

\subsection{Miniband conduction (MBC)}
For zero electric field Eq.~(\ref{Eqhamz}) is diagonalized by
a set of Bloch functions $\varphi_q(z)=\sum_n e^{inqd}\Psi_n(z)$
(here the wave function $\Psi_n(z)$ is localized in quantum well
$n$, e.g., a Wannier function)
and the dispersion relation is given by the miniband
$E(q)=2T_1\cos(qd)$.
The  stationary Boltzmann equation for the distribution function
$f(q,{\bf k})$ is then
\begin{equation}
\frac{eF}{\hbar}\frac{\partial f(q,{\bf k})}{\partial q}=
\frac{n_F(E(q)+E_k)-f(q,{\bf k})}{\tau(E(q)+E_k)}
\label{Eqboltz}\end{equation}
where  the relaxation-time is allowed to depend on energy,
see, e.g., Ref.\cite{prl1} for a suitable model.  Once the
solution to 
Eq.~(\ref{Eqboltz}) is found, the current
is calculated from
\begin{equation}
J(F)=\frac{e}{4\pi^3\hbar}\int {\rm d}^2 k\int_{-\pi/d}^{\pi/d}{\rm d}q
f(q,{\bf k})\frac{{\rm d}E(q)}{{\rm d}q}\, .
\label{JBoltz}
\end{equation}
The electron density per period is given by
\begin{equation}
N_{2D}=\frac{d}{4\pi^3}\int {\rm d}^2 k\int_{-\pi/d}^{\pi/d}{\rm d}q
f(q,{\bf k})
\end{equation}
and is used to determine the chemical potential for a given
electron density. 
This approach can be extended beyond the relaxation time
approximation\cite{IGN91,LEI91}, but the generic features
remain unchanged.

\subsection{Wannier-Stark hopping (WSH)}
In the presence of an
electric field, the eigenstates of the Hamiltonian
become the localized Wannier-Stark states,
\begin{equation}
\phi_{\nu}(z)=\sum_n J_{n-\nu}\left(\frac{2T_1}{eFd}\right)
\Psi_n(z)\label{EqWSstate}
\end{equation}
with energy $E_{\nu}=-\nu eFd$, and $J_n(z)$ is the Bessel function of the
first kind. Scattering causes hopping between the different states.
Within Fermi's golden rule, the current is given by
\begin{eqnarray}
J(F)&=&\sum_{l>0} l\frac{e}{\tau_0}  
\sum_n\left[J_{n}\left(\frac{2T_1}{eFd}\right)
J_{n-l}\left(\frac{2T_1}{eFd}\right)\right]^2 \nonumber\\
&\quad&\times
\frac{1}{2\pi^2}\int{\rm d}^2 k \left[n_F(E_k)-n_F(E_k+leFd)\right]\, .
\end{eqnarray}
Here the term $\sum [J_{n}J_{n-l}]^2$ arises due to the spatial overlap 
of the Wannier-Stark functions.
The electron density per period is given by:
\begin{equation}
N_{2D}=\rho_0k_BT_e\log\left[1+\exp\left(\frac{\mu}{k_BT_e}\right)\right]
\end{equation}
which relates $\mu$ to $N_{2D}$.
Again, it is possible to generalize this approach to
more realistic scattering 
mechanisms \cite{ROT97,BRY97}.
\begin{figure}
\begin{center}
\begin{tabular}{|c|c|c|c|} 
\hline
 & coupling $ T_1$ & field drop $eFd$ &
scattering $\Gamma$\\ \hline &&&\\
\begin{minipage}{3cm}
Miniband conduction\\ 
\epsfig{file=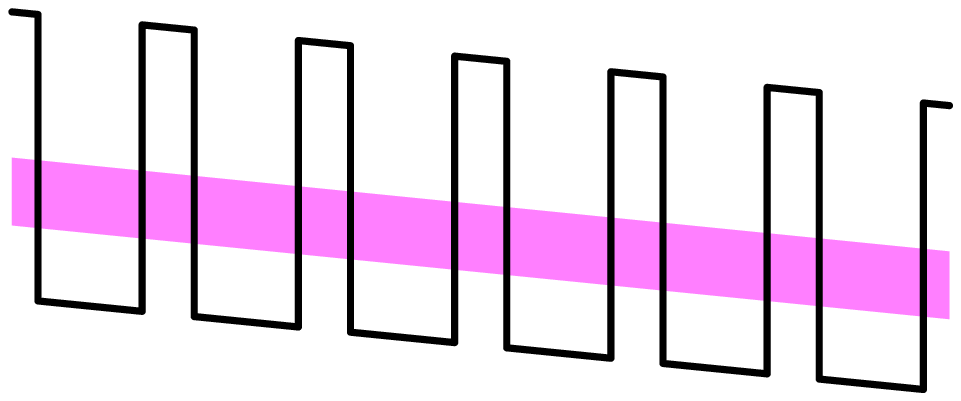,width=3cm}
\end{minipage}
&\begin{minipage}{2.0cm} exact\\
miniband\end{minipage} &
acceleration &
golden rule \\ \hline & \multicolumn{2}{|c|}{ }&\\
\begin{minipage}{3cm}
Wannier Stark\\ hopping\\
\epsfig{file=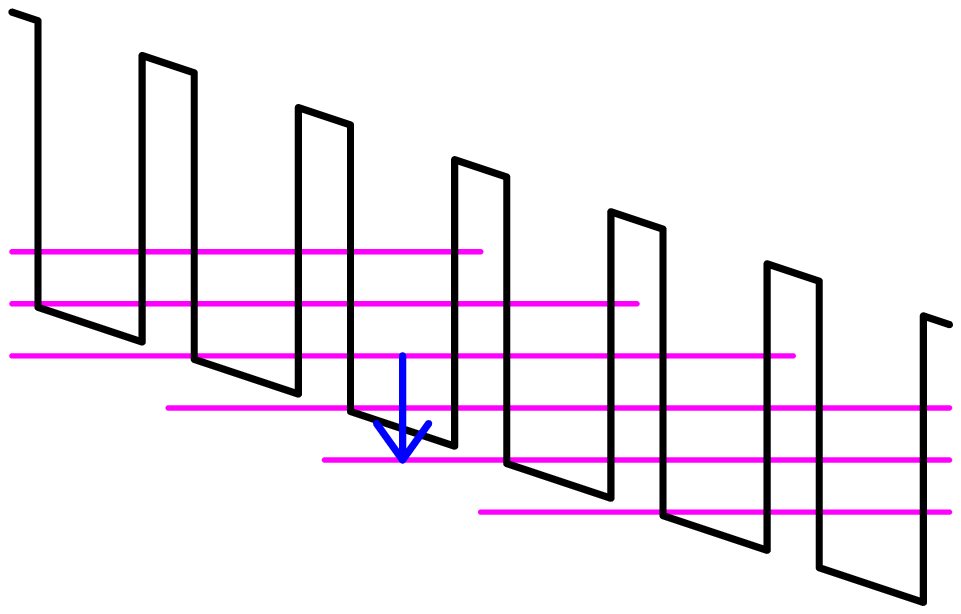,width=3cm}\end{minipage} &
 \multicolumn{2}{|c|}  {exact: Wannier Stark states} &
golden rule \\ \hline  &&&\\
\begin{minipage}{3cm}Sequential tunneling\\
\epsfig{file=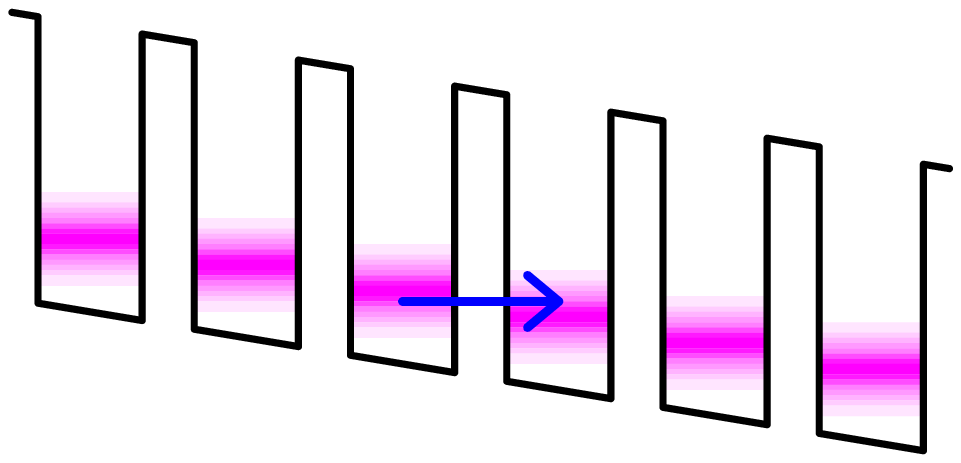,width=3cm}\end{minipage} &
 lowest order& \begin{minipage}{2.0cm} energy mismatch\end{minipage} & 
\begin{minipage}{2.0cm} "exact"\\ spectral function
\end{minipage} \\ \hline
\end{tabular}
\caption[]{Overview of standard theoretical models for superlattice
transport}
\label{Fig1}
\end{center}
\end{figure}

\subsection{Sequential tunneling (ST)}
In this approximation the phase information is
lost after each tunneling event between adjacent wells.
The scattering within a well
is treated self-consistently by 
solving for the spectral functions
$A({\cal E},{\bf k})$; in this work we use the
self-consistent Born-approximation\cite{MAH90} 
for the self-energy.
The transitions to neighboring wells are
calculated in lowest order of the coupling yielding \cite{WAC97,MAH90,MUR95}:
\begin{equation}
J(F)=\frac{e}{2\pi^2}\int{\rm d}^2 k 
\int\frac{{\rm d}{\cal E}}{2\pi\hbar } T_1^2 A({\cal E},{\bf k})
A({\cal E}+eFd,{\bf k})
  \left[n_F({\cal E})-n_F({\cal E}+eFd)\right]\;.
\end{equation}
The carrier density is given by:
\begin{equation}
N_{2D}=\frac{1}{2\pi^2}\int{\rm d}^2 k
\int\frac{{\rm d}{\cal E}}{2\pi} n_F({\cal E})A({\cal E},{\bf k})\, .
\end{equation}
This approach gives quantitative agreement with experiments
in weakly coupled structures
when realistic models for impurity and interface
scattering are employed \cite{WAC97b,WAC97}.

\subsection{Summary}
The important issue to recognize is that these three approaches 
treat scattering, external field, and coupling within
different approximations. 
MBC does not properly include field-induced 
localization because of its inherent 
assumption of extended states, WSH treats scattering in lowest order 
perturbation theory (in particular, there is no 
broadening of the states), and ST is explicitly lowest order in 
the interwell coupling.
The basic features of these models are summarized in Figure 1.

\subsection{Nonequilibrium Green functions (NGF)} 
The basic building blocks are the correlation and retarded
Green functions:
\begin{eqnarray}
G^<_{m,n}(t,t',{\bf k})&=&
i\langle a_{n}^{\dag}(t',{\bf k})a_{m}(t,{\bf k})\rangle \\
G^{\rm ret}_{m,n}(t,t',{\bf k})&=&-i\Theta(t-t')
\langle \{a_{m}(t,{\bf k}),a_{n}^{\dag}(t',{\bf k})\}\rangle ,
\end{eqnarray}
where  
and $a_n^{\dag}(t,{\bf k})$ and $a_n(t,{\bf k})$ are the creation
and annihilation operators for the state
$\Psi_n(z)e^{i({\bf k}\cdot {\bf r})}/A$ in well $n$.
These functions obey the Dyson and Keldysh equations, respectively,
given below for the superlattice case.
Ref.\cite{HAU96} may be consulted for a text-book discussion.
The observables, such as
the momentum distribution, current, and electron density are
computed from $G^<_{m,n}(t,t',{\bf k})$:
\begin{eqnarray} 
f_{\rm QM}(q,{\bf k})&=&\ 
\int {d{\cal E}\over 2\pi i}
\sum_n e^{-ihqd}G^<_{n,0}({\cal E},{\bf k})\label{Momdistr}\\ 
J(F)&=&\frac{e}{2\pi^2}\int {d{\cal E}\over 2\pi}
\int{\rm d}^2 k \frac{2}{\hbar}\,{\rm Re}
\left\{T_1 G^<_{n+1,n}({\cal E},{\bf k})\right\}\label{Eqneqstrom}\\
N_{2D}&=&\frac{1}{2\pi^2}\int {d{\cal E}\over 2\pi i}\int{\rm d}^2 k\, 
G^<_{n,n}({\cal E},{\bf k})\label{Eqneqdichte}
\end{eqnarray}
These expressions exploit the fact that
in the stationary state the Green functions only depend
on the time difference $\tau=t-t'$, and one can define a
Fourier transformation via \cite{HAU96}:
\begin{equation}
G_{m,n}({\cal E},{\bf k})=\int {\rm d}\tau 
e^{i\left({\cal E}-eFd\frac{n+m}{2}\right)\frac{\tau}{\hbar}}
G_{m,n}(t,t-\tau,{\bf k})\, .
\end{equation}
Without scattering between the {\bf k}-states and at $T_1=0$
the Green-functions are diagonal in the well index: 
$G^{\rm ret}_{m,n}({\cal E},{\bf k})=
\delta_{m,n}g^{\rm ret}_{n}({\cal E},{\bf k})$
with the free particle Green-function 
$g^{\rm ret}_{n}({\cal E},{\bf k})=1/({\cal E}-E_k+i0^+)$.
The full Green function is then determined by the Dyson equation:
\begin{eqnarray}
G_{m,n}^{\rm ret}&(&{\cal E},{\bf k})=
g^{\rm ret}_{m}\left({\cal E},{\bf k}\right)+
g^{\rm ret}_{m}\left({\cal E}+eFd\frac{m-n}{2},{\bf k}\right)\nonumber \\
&&\times \sum_l\Sigma_{m,l}^{\rm ret}
\left({\cal E}+eFd\frac{l-n}{2},{\bf k}\right)
G_{l,n}^{\rm ret}\left({\cal E}+eFd\frac{l-m}{2},{\bf k}\right)\, .
\label{Eqdyson}
\end{eqnarray}
The self-energy will be written as
\begin{equation}
\Sigma_{m,n}^{\rm ret}({\cal E},{\bf k})=
\delta_{m,n}\tilde{\Sigma}^{\rm ret}_n({\cal E},{\bf k})
+T_1\delta_{m+1,n}+T_1\delta_{m-1,n}\label{Eqsigma}
\end{equation}
where $\tilde{\Sigma}$ contains contributions both from impurity
and phonon scattering.  The relevant expressions are
\begin{eqnarray} 
\tilde{\Sigma}^{{\rm ret}/<}_{n,{\rm imp}}({\cal E})&=& 
\frac{N_d}{A}\sum_{{\bf k}'}V_{\rm imp}^2 
G_{n,n}^{\rm ret/<}({\cal E},{\bf k}')\label{EqSigimp} 
\end{eqnarray} 
for impurities, while for optical phonon scattering  
we take (see, e.g., Ch. 4.3 of Ref.~\cite{HAU96} 
for the derivation):  
\begin{eqnarray} 
\tilde{\Sigma}^{<}_{n,{\rm o}}({\cal E}) &=& 
\frac{|M_{\rm o}|^2}{A}\sum_{{\bf k}'}  
\Big\{N_{\rm o}G^{<}_{n,n}({\cal E}-\hbar\omega_{\rm o},{\bf k}')\nonumber\\ 
&\quad&\quad
+(N_{\rm o}+1)G^{<}_{n,n}({\cal E}+\hbar\omega_{\rm o},{\bf k}')\Big\} 
\label{EqSiglessopt}
\end{eqnarray}
\begin{eqnarray}
\tilde{\Sigma}^{\rm ret}_{n,{\rm o}}&(&{\cal E})= 
\frac{|M_{\rm o}|^2}{A}\sum_{{\bf k}'} \Big\{ 
(N_{\rm o}+1)G^{\rm ret}_{n,n}({\cal E}-\hbar\omega_{\rm o},{\bf k}') 
+N_{\rm o}G^{\rm ret}_{n,n}({\cal E}+\hbar\omega_{\rm o},{\bf k}')\nonumber \\ 
&&
+i\int\frac{d{\cal E}'}{2\pi}  
G^{<}_{n,n}({\cal E}-{\cal E}',{\bf k}') 
\left[\frac{1}{{\cal E}'-\hbar\omega_{\rm o}+i0^+}- 
\frac{1}{{\cal E}'+\hbar\omega_{\rm o}+i0^+}\right]\Big\}\, . 
\label{EqSigretopt} 
\end{eqnarray} 
It is possible to simulate acoustic phonons with a similar expression,
using a small fictitious discrete energy $\hbar\omega_{ac}$ \cite{prl2}.
In the above, we also gave the self-energy expressions needed
in the Keldysh equation:
\begin{eqnarray}
G^{<}_{m,n}({\cal E},{\bf k})&=&\sum_{m_1}
G^{\rm ret}_{m,m_1}\left({\cal E}+eFd\frac{m_1-n}{2},{\bf k}\right)
G^{\rm adv}_{m_1,n}\left({\cal E}+eFd\frac{m_1-m}{2},{\bf k}\right)
\nonumber \\ 
&&\times \tilde{\Sigma}^{<}_{m_1}
\left({\cal E}+eFd\left(m_1-\frac{m+n}{2}\right),{\bf k} \right) \, .
\label{EqGlessom}
\end{eqnarray}
\begin{figure}
\begin{center}
\epsfig{file=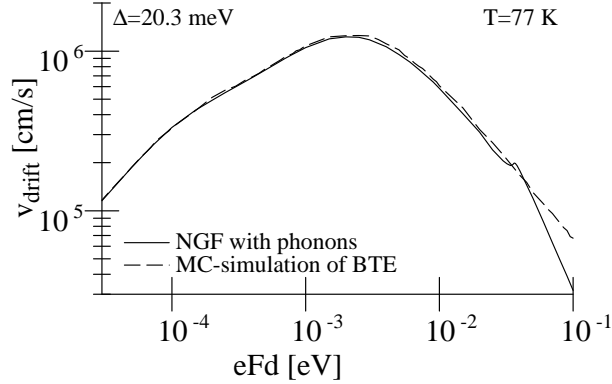,width=8cm}
\caption{Drift velocity for a wide-band superlattice.  Full line:
Calculation by nonequilibrium Green functions.  Dashed line:
Monte Carlo simulation of Boltzmann's transport equation 
(from Ref.[12]).}
\label{Fig2}
\end{center}
\end{figure}
The numerical evaluation of these equations has been discussed in
two recent publications \cite{prl1,prl2}, and here we just summarize
our basic strategy, and give a few representative results.  
The first step in the analysis consists of evaluating the 
current-voltage characteristics for the different
models, and an example of such a calculation is given in Fig. 2.
We note that the Boltzmann equation gives an excellent agreement
with the full quantum result, in particular for low electric
fields.  The additional structure in the quantum mechanical
curve is a phonon-replica: the Boltzmann equation cannot capture
features like this because the electric field does not enter
as an energy-scale in the collision integral. 
It is also of interest to note that the use of a realistic
electron-phonon scattering model, instead of a simple relaxation
time, leads to a current-voltage characteristic which deviates
significantly from the simple Esaki-Tsu shape,
$v_{\rm drift}\propto F/(F^2 + F_{\rm crit}^2)$.  By performing a large
number of calculations like the one depicted in Figure 2 it is possible
to develop visual criteria as to the validity of the various simplified
approaches discussed in Section 2.  These phenomenological  considerations
can be made more rigorous by examining in detail the retarded Green
function, which determines the quantum mechanical correlation between
quantum wells $n$ and $m$ \cite{prl1}.  For example, if the terms
$G^{\rm ret}_{n\pm 1,n}$ are of the order of $G^{\rm ret}_{n,n}$, the
wave-function is delocalized.  By using the expressions given in 
Ref.\cite{prl1}, one finds that the states are delocalized if
$2|T_1|\gg\Gamma$ and $2|T_1|\gg eFd$, i.e., the Boltzmann miniband
picture is a useful starting points.  Similarly, by demanding that
the $G^{\rm ret}_{m,n}$ vanish if $m\ne n$, one can find the regime
where sequential tunneling dominates.  The respective ranges of
validity can be presented as a ``phase digram'', Fig. 3.
\begin{figure}
\begin{center}
\epsfig{file=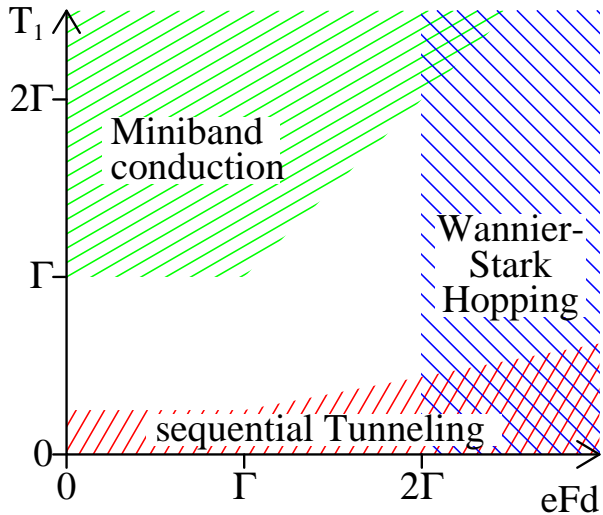,width=8cm}
\caption{Ranges of validity for different transport models}
\end{center}
\label{Fig3}
\end{figure}
 
Additional insight to the differences between quantum kinetics
and Boltzmann kinetics can be obtained by examining the momentum
distribution functions.  Figure 4 presents such a comparison.
We direct attention to the following features.  At low fields,
Fig. 4(a), the distribution function can be viewed as a distorted thermal
equilibrium function, and linear response theory holds, corresponding
to the linear part of the current-voltage characteristics of Fig. 2 
(however, at very low temperatures one should consider weak
localization effects, not included in the present choice of
the impurity self-energy, and the agreement between quantum
and Boltzmann calculations may be weakened). If the electric
field increases, electron heating becomes important.  For moderate
fields the distribution function resembles a distorted equilibrium
function, but with an elevated electron temperature, Fig. 4(b).
At even higher fields, close to the maximum in the IV-curve,
the distribution function strongly deviates from any kind
of equilibrium in $q$ space, Fig. 2(c).  The results for quantum
and semiclassical calculations look similar in the negative differential
conductance (NDC) regime (not shown in Fig. 4.).  The situation 
changes dramatically when the field energy is larger than the
mini-band width, $edF\ge \Delta$, Fig. 4(d).  As the electrons can perform 
several Bloch oscillations in the semiclassical picture, the distribution function
is almost flat withing the Brillouin zone of the miniband.  The latter
holds for the NGF result as well.  However, the absolute values of the 
distribution functions differ 
significantly.  
\begin{figure} 
\begin{center}
\noindent\epsfig{file=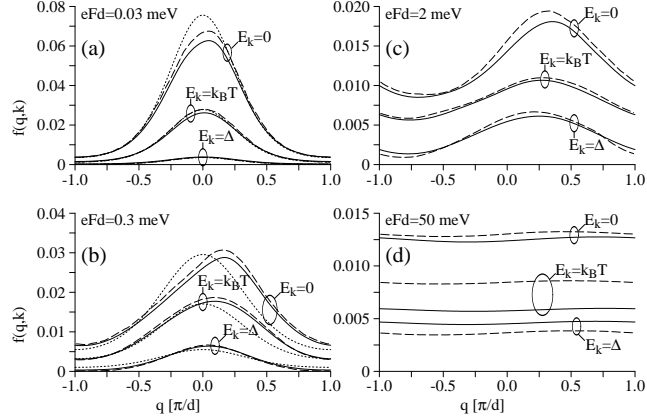,width=8.5cm}\\[0.2cm] 
\caption[a]{Electron distribution versus quasimomentum in the 
Brillouin zone of the miniband  for different values 
of $k$ (from Ref.[12]).  
The parameters as Fig.~\ref{Fig2}. 
Full line: NGF calculation. Dashed line:  
MC-simulation of BTE. The dotted line shows the thermal 
distribution $\propto \exp[-E(q,{\bf k})/k_BT_e]$ with $T_e=T$ in (a)
and $T_e=140$ K in (b) for comparison.
\label{Fig4}} 
\end{center}
\end{figure} 
The reason is the modification in scattering
processes due to the presence of the electric field, leading to
significant deviations in the distribution function.  We noticed
this phenomenon already in our discussion of the current-voltage
characteristics of Fig. 2.  The strong changes in the quantum
momentum distribution are further illustrated in Fig. 5.
\begin{figure} 
\begin{center}
\epsfig{file=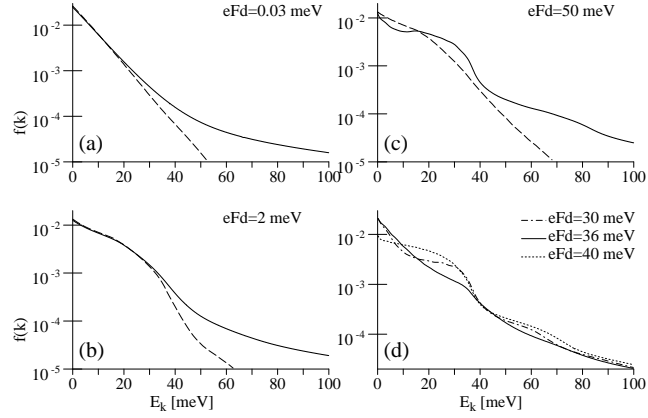,width=8.5cm}\\[0.2cm] 
\caption[a]{Electron distribution versus quasimomentum $k$ 
(from Ref.[12]).  (a,b,c): Comparison  
between NGF calculation (full line) and BTE (dashed line); 
(d): Results from NGF for different fields. The BTE result (not shown) 
resembles the result from (c) for all three fields. 
\label{Fig5}}
\end{center} 
\end{figure} 

We can summarize the results given in the first part of this paper
as follows.  In wide-band superlattices the Boltzmann equation gives reliable 
results concerning linear response at low fields, 
electron heating at moderate fields, and the onset of negative 
differential conductivity.  
In contrast, for high electric fields or weakly coupled 
SLs significant differences appear. 
In this case the quantum nature of transport is important and  
a semiclassical calculation may be seriously in error. 

\section{Superlattice as a THz-photon detector}
\subsection{Introduction}
     It  has  recently  been estimated \cite{ref35}  that  the  room
temperature   current  responsivity  of  a   superlattice
detector  ideally coupled to the THz-photons  can  nearly
reach  the  quantum efficiency  $e/\hbar\omega$
 in the limit of high frequencies
$\omega\gg\nu$ (here $\omega$  is  the  incident
radiation  frequency and $\nu$ is a characteristic
scattering frequency).
This   value  of  the  responsivity  is  being   normally
considered  as  a  quantum limit for detectors  based  on
superconducting   tunnel  junctions  operating   at   low
temperatures \cite{ref18}. For high frequencies the mechanism of the
THz-photons detection in superlattices was described \cite{ref35} as
a   bulk   superlattice   effect  caused   by   dynamical
localisation of electrons.

Here we describe a recent theory of the superlattice current 
responsivity \cite{JAP}. We focus on relative broad-band superlattices
($\Delta\simeq 20$ meV), and field strengths smaller than the
onset of negative differential conductivity.  Thus, 
according to the analysis
given above, a Boltzmann equation based description should suffice.
One should note, however, that here we investigate a time-dependent
situation, and that a microscopic analysis in the spirit of the
first part of this paper has not, to our knowledge, been
carried out.
We use an equivalent circuit for the superlattice coupled to
a broadband antenna (see Fig. 6), which is similar to the
equivalent circuit used in resonant tunneling \cite{ref37} and
Schottky diode \cite{ref38} simulations. The suggested  equivalent
circuit of the device allows one to treat microscopically
the  high-frequency  response of the  miniband  electrons
and,  simultaneously, take into account a finite matching
efficiency   between  the  detector   antenna   and   the
superlattice  in  the presence of parasitic  losses.  Our
analytic  results  lead  to  the  identification  of   an
important  physical  concept: the  excitation  of  hybrid
plasma-Bloch \cite{ref39}  oscillations 
{\it in the  region  of  positive
differential   conductance  of  the   superlattice}.   
Numerical  computations \cite{JAP}, performed for  room  temperature
behavior  of  currently available  superlattice  diodes,
show   that   both  the  magnitudes  and   the   roll-off
frequencies  of the responsivity are strongly  influenced
by  this  effect.  The  excitation  of  the  plasma-Bloch
oscillations gives rise to a resonant-like dependence  of
the  responsivity  on  the incident radiation  frequency,
improving essentially the coupling of the superlattice to
the detector antenna. We will also show that peak current
densities  in  the device and its geometrical  dimensions
should  be  properly optimized in order  to  get  maximum
responsivity for each frequency of the incident  photons.
Finally,  we  will  present numerical  estimates  of  the
responsivity for the 1-4 THz frequency band  and  compare
its  value  with  the  quantum efficiency  $e/\hbar\omega$ 
of  an  ideal
detector.
\begin{figure}
\begin{center}
\epsfig{file=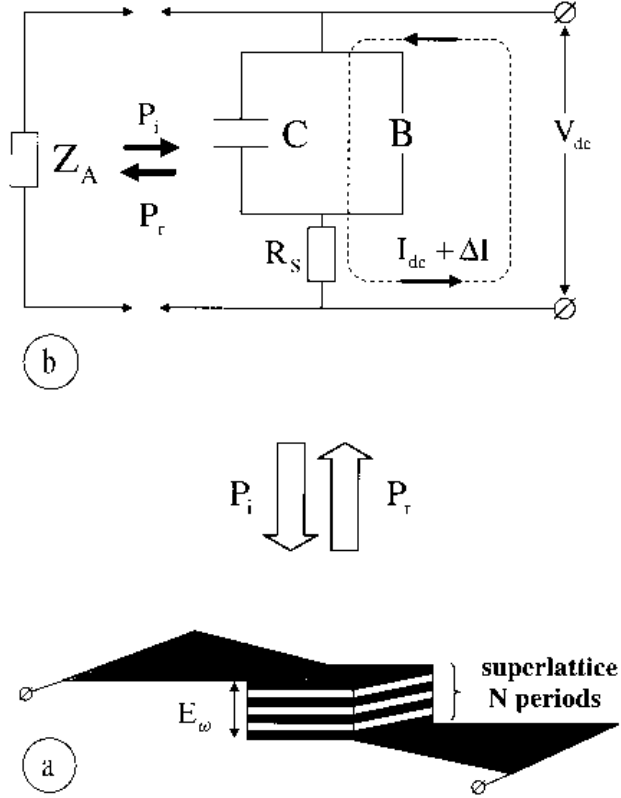,width=8.5cm}\\[0.2cm]
\caption{(a)  THz-radiation  coupled  to  a  $N$-period
semiconductor superlattice by a co-planar broad band bow-
tie  antenna,  $P_i$ and $P_r$  are the incident and reflected powers
respectively.  (b) Equivalent circuit  for  a  THz-photon
detector  with  a  dc  voltage  bias  source:  $B$--miniband
electrons  capable  to  perform  Bloch  oscillations,  
$C$--superlattice capacitance, $R_S$--parasitic series  resistance,
$Z_A$--bow-tie antenna impedance, $V_{dc}$--dc bias voltage.}
\label{Fig6}
\end{center}
\end{figure}
\subsection{Theoretical formalism}
The exact solution of the time-dependent Boltzmann equation
in the relaxation time approximation for an arbitrary
time-dependent  electric field  can be presented  in  the
form of a path integral \cite{ref41}:
\begin{equation}
f(q,{\bf k},t) = \int_{-\infty}^t \nu dt_1
\exp\left[ - \nu (t - t_1)\right ]
f_0(q - \int_{t_1}^t e/\hbar F(t_2) dt_2,{\bf k})\;.
\label{(10)}
\end{equation}
Using  Eqs.(\ref{JBoltz})  and  (\ref{(10)})  we find  the  time-dependent
current  $I(t)$ describing ac transport in a superlattice,  with
electron  performing  ballistic  motion  in  a  mini-band
according  to  the  acceleration  theorem  and  suffering
scattering \cite{ref2,ref3}:
\begin{equation}
I(t) = 2I_p \int_{-\infty}^t \nu dt_1
\exp\left[ -\nu (t-t_1)\right]
\sin\left[ {e\over N\hbar} \int_{t_1}^t V(t_2)dt_2\right ]\\;,
\label{(11)}
\end{equation}
where $V(t) = LF(t)$ is   the   voltage   across   the   superlattice
perpendicular to the layers,  $L = N d$ is the superlattice length,
$N$ is  the  number of periods in the superlattice sample, 
$I_p = S j_p$, $S = \pi a^2$
is  the  area  of the superlattice,  $a$ is the  superlattice
mesa radius, and
\begin{equation}
j_p = e {v_0\over 2} \int {2d{\bf k}dq\over (2\pi\hbar)^3} 
\cos \left(q d\right)
f_0(q,{\bf k}) 
\label{(12)}
\end{equation}
is  the  characteristic current density. The  integration
over $q$  in Eq.(\ref{(12)}) must be carried out over the Brillouin
zone  $-\pi/d\le q \le \pi/d$.

      The   peak  current  density  $j_P$ and  the  scattering
frequency  $\nu$ can  be considered as the main parameters  of
the  employed  model. They can readily be estimated  from
experimentally  measured or numerically simulated  values
of $I_P$  and $V_P$. For both degenerate and non-degenerate electron
gas one gets \cite{ref2,ref3}
\begin{equation}
j_P = en {v_0\over 2}
\label{(14)}
\end{equation}
if  $\Delta\gg kT,\epsilon_F$,
where $kT$ is the equilibrium thermal excitation energy,
$\epsilon_F = \hbar^2(3 \pi^2 N_{2D})^{2/3}/(2m_{\rm eff}) $
is  the  Fermi  energy of degenerate electrons,    
$m_{\rm eff} = m_{zz}^{1/3} m^{2/3}$
is  the  density  of  states
effective  mass  near the miniband bottom,  and 
$m_{zz} = 2\hbar^2 /\Delta d^2$  is  the
effective mass of electrons along the superlattice  axis.
In  the  particular  case  of the  Boltzmann  equilibrium
distribution function Eq.(\ref{(12)}) yields \cite{ref3}
$j_P=(env_0/2) [I_1(\Delta/2kT)/I_0(\Delta/2kT)]$, 
where $I_{0,1}$  are  the
modified Bessel functions.

We now suppose that in addition to the dc voltage  $V_{SL}$,
an   alternating  sinusoidal  voltage  with   a   complex
amplitude $V_\omega$ is applied to the superlattice:
\begin{equation}
V(t) = V_{SL} + {1\over 2}
\left[V_\omega \exp(i\omega t) +
V_\omega^* \exp(-i\omega t)\right]\;.
\label{(15)}
\end{equation}
Generally,  $V_{SL}, V_\omega$
can  be  found  from  an  analysis  of   the
equivalent circuit given in
Fig.\ \ref{Fig6}. We write the ac voltage amplitude as 
$V_\omega = |V_\omega| e^{i\psi}$; 
both $|V_\omega|$   and $\psi$
can  be  obtained  self-consistently  taking  account  of
reflection  of the THz photons from the superlattice  and
their absorption in the series resistor $R_S$.

    Making use Eq.(\ref{(11)}) we obtain \cite{ref3}:
\begin{equation}
I(t) = 2I_P \int_0^{\infty} \nu dt_1 \exp(-\nu t_1)
\sin\left[ {eV_{SL}\over N\hbar} t_1 + \Phi(t,t_1) \right ]\\;,
\label{(16)}
\end{equation}
where
\begin{equation}
\Phi(t,t_1) = {e\over N\hbar\omega} \times
{1\over 2} \left\{ iV_\omega\exp(i\omega t)
[\exp(-i\omega t_1)-1] + c.c. \right \}\;.
\label{(17)}
\end{equation}
According  to  Eq.(\ref{(16)}),  electrons  in  a  superlattice
miniband  perform  damped  Bloch  oscillations  with  the
frequency
$\Omega_B = eV_{SL} / N\hbar = eE_{SL} d/\hbar$, 
and the phase $\Phi(t,t_1)$ modulated by the external  ac
voltage.

      Equation  (\ref{(16)})  contains,  as  special  cases   the
following  results:  (i) a harmonic   voltage   $V(t)$
($V_{SL}=0$) leads  to
dynamical  localisation, and current harmonics generation
with  oscillating power dependence \cite{ref3}; (ii) a  dc  current-
voltage  characteristics  of the irradiated  superlattice
$I_{DC}(V_{SL},V_\omega) = (\omega/2\pi) \int I(t) dt$
shows  resonance  features (`Shapiro steps')  leading  to
absolute   negative  conductance \cite{ref3,ref5,ref7,ref17};  
(iii)   and   to
generation  of dc voltages (per one superlattice  period)
that are multiples of $\hbar\omega/e$\cite{ref13}.

\subsection{Current responsivity}
      
We   define  the  current  responsivity\cite{ref18}  of   the
superlattice detector as the current change $\Delta I$  induced  in
the external dc circuit per incoming ac signal power $P_i$:
\begin{equation}
R_i(\omega,V_{SL})= {\Delta I \over P_i}\;.
\label{(31)}
\end{equation}
This  definition  takes into account both  the  parasitic
losses  in  the  detector and the finite  efficiency  for
impedance-matching  of  the  incoming  signal  into   the
superlattice  diode.

It can be shown \cite{JAP}, that in the small-signal
approximation both the dc current change 
$\Delta I^{SL}_{DC}$ and  the  power $P^{SL}_{\rm abs}$
absorbed  in  the  superlattice are proportional  to  the
square modulus of the complex voltage $|V_\omega|^2$. 
This circumstance
permits  us  to  calculate  $|V_\omega|^2$  self-consistently for  given
values  of  the incoming power,  making use  a  linear  ac
equivalent  circuit analysis and, then, find the  current
responsivity $R_i(\omega,V_{SL})$.

     The  results  of the calculation of the superlattice
current  responsivity   $R_i(\omega,V_{SL})$
are presented  in  the  following
form:
\begin{equation}
R_i(\omega,V_{SL})=
{R_i^{(0)}(\omega,V_{SL}) A(\omega,V_{SL}) \over
1+ R_S (dI^{SL}_{DC}(V_{SL})/dV_{SL}) }\;,
\label{(32)}
\end{equation}
where
\begin{equation}
R_i^{(0)}(\omega,V_{SL}) = -{e\over N \hbar\nu}
{(V_{SL}/V_P)[3 + (\omega\tau)^2 - (V_{SL}/V_P)^2]\over
[1+(V_{SL}/V_P)^2]
[1+(\omega\tau)^2-(V_{SL}/V_P)^2]}
\label{(33)}
\end{equation}
is the superlattice current responsivity under conditions
of  a  perfect  matching and neglecting parasitic  losses,
$R_S\to 0$. \cite{ref35}

     The factor $A(\omega,V_{SL})$  
in Eq. (\ref{(32)}) describes the effect of  the
electrodynamical  mismatch between the  antenna  and  the
superlattice  and  the signal absorption  in  the  series
resistance
\begin{equation}
A(\omega,V_{SL})=
\left[1 - 
\left|{
Z_A-(Z^{SL}_{AC}(\omega,V_{SL}) + R_S) \over
Z_A+(Z^{SL}_{AC}(\omega,V_{SL}) + R_S)}
\right|^2
\right]\times
{{\rm Re}Z^{SL}_{AC}(\omega,V_{SL})\over
{\rm Re}Z^{SL}_{AC}(\omega,V_{SL})+R_S}\;.
\label{(34)}
\end{equation}
The first factor in Eq. (\ref{(34)}) describes the reflection  of
the  THz-photons due to mismatch of the antenna impedance $Z_A$
and  the total impedance of the device
$Z^{SL}_{AC}(\omega,V_{SL})+R_S$, with the second
one  being responsible for sharing of the absorbed  power
between  the active part of the device described  by  the
impedance $Z^{SL}_{AC}(\omega,V_{SL})$ and the series resistance $R_S$.

    The superlattice impedance is defined as
\begin{equation}
Z^{SL}_{AC}(\omega,V_{SL})=
1/\left[G^{SL}_{AC}(\omega,V_{SL}) + i\omega C\right]\;,
\label{(35)}
\end{equation}
where $G^{SL}_{AC}(\omega,V_{SL})$ is the superlattice conductance,
$C=\epsilon_0 S / 4\pi L$  is the capacitance
of  the  superlattice,  and  
$\epsilon_0$ is  the  average  dielectric
lattice constant.

     Finally,  the last factor in 
the denominator of Eq. (\ref{(32)}) describes  the
redistribution of the external bias voltage  $V_{DC}$
between  the
dc  differential resistance of the superlattice  
$(dI^{SL}_{DC}(V_{SL})/dV_{SL})^{-1}$ and  the
series  resistance  $R_S$, with the dc voltage  drop  on  the
superlattice  $V_{SL}$ being determined by the  solution  of  the
well-known load equation \cite{ref18}
\begin{equation}
V_{DC}=V_{SL} + I^{SL}_{DC}(V_{SL})R_S\;.
\label{(36)}
\end{equation}

\subsection{Superlattice  dielectric function.  Hybridisation  of
Bloch and plasma oscillations}

    We next analyze the condition of optimized matching of
the superlattice to the incident radiation.  Assuming the limit of  negligible
series  resistance $R_S\to 0$ this condition can be  obtained  from
the solution of the equation
\begin{equation}
Z^{SL}_{AC}(\omega,V_{SL})=Z_A
\label{(41)}
\end{equation}
for  the complex frequency
$\omega(V_{SL})$ . This solution determines the
resonant  line position and the line width at  which  the
absorption  in  the  superlattice tends  to  its  maximum
value.

One can transform 
Eq. (\ref{(41)})
to the following form:
\begin{equation}
\epsilon(\omega,E_{SL}) = 
{\epsilon_0\over i\omega C Z_A}\;,
\label{(42)}
\end{equation}
where
\begin{equation}
\epsilon(\omega,E_{SL}) = \epsilon_0
+ {4\pi \sigma_0\over i \omega}
F_1(\omega,E_{SL})
\label{(43)}
\end{equation}
is  the dielectric function of the superlattice, with the
dc  field  $E_{SL}$ being applied to the device \cite{ref38},  
and $F_1(\omega,E_{SL})$ is defined  by
\begin{equation}
F_1(\omega,V_{SL}) = 
{1 + i\omega\tau - (V_{SL}/V_P)^2 
\over
\left [1+(V_{SL}/V_P)^2 \right]
\left[(1+i\omega\tau)^2 + (V_{SL}/V_P)^2\right] }\;.
\label{(21)}
\end{equation}
\begin{figure}
\begin{center}
\epsfxsize=70mm
\epsffile{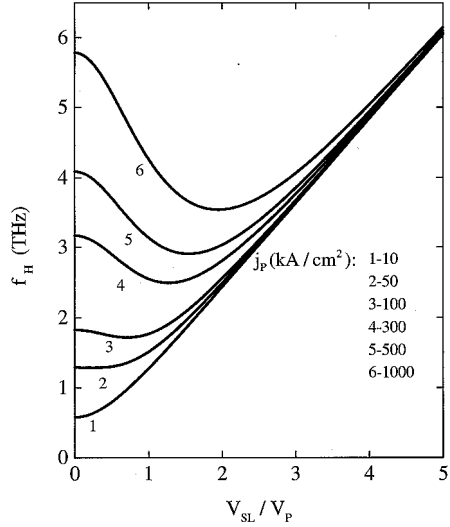}
\caption{
The calculated hybrid plasma-Bloch oscillation
frequency  $f_H$ as a function of  the  normalized
superlattice  voltage drop $V_{SL}/V_P$ for different values  of  the
peak  current densities $j_P=$ 10, 50, 100, 300, 500,  and  1000
kA/cm$^2$ (from Ref.[24]). Typical values of the superlattice parameters ( $d$
= 50{\AA},  $E_P$=10 kV/cm, $\epsilon_0$ =13) were used for the calculations.
}
\label{Fig7}
\end{center}
\end{figure}
    In the high-frequency limit  
$\epsilon_0/CZ_A\omega \to 0$ the solution of Eq. (\ref{(41)})
coincides with the solution of the equation
\begin{equation}
\epsilon(\omega,E_{SL})=0
\label{(44)}
\end{equation}
describing the eigenfrequencies  $\omega^H_\pm$
of the hybrid  plasma-Bloch
oscillations in a superlattice, \cite{ref38}
\begin{equation}
\omega^H_\pm(E_{SL}) = \pm \omega_P
\left [ {1\over 1 + (E_{SL}/E_P)^2}
+ \left( {\nu\over\omega_P}\right)^2
(E_{SL}/E_P)^2 \right]^{1/2} + i\nu
\label{(45)}
\end{equation}
where  $\omega_P$  is  the  plasma  frequency  of  electrons  in   a
superlattice. The plasma frequency $\omega_P$ 
can be given in  terms
of  the small-field dc conductivity $\sigma_0$ or, equivalently, in
terms of the peak current density $j_P$
\begin{equation}
\omega_P =
\left({4\pi\sigma_0 \nu\over\epsilon_0}\right)^{1/2} =
\left({8\pi j_P ed\over\epsilon_0\hbar}\right)^{1/2}/;.
\label{(46)}
\end{equation}
Equation  (\ref{(46)})  reduces in the particular case  of 
wide-miniband superlattices ($\Delta\gg kT,\epsilon_F$) 
to the standard formula 
$\omega_P = (4\pi e^2 n / \epsilon_0 m_{zz})^{1/2}$.

     In  the  limiting case of small applied dc  electric
fields  $E_{SL}/E_P \to 0$ one  finds from 
Eq. (\ref{(45)}) the plasma frequency
$\omega^H_\pm \to \pm\omega_P$,
while  in  the  opposite case 
$E_{SL}/E_P \to \infty$, the Bloch  frequency 
$\omega^H_\pm\to\pm\Omega_B=\pm eE_{SL} d/\hbar$  is
recovered.  The  scattering frequency  $\nu$ in  Eq.  (\ref{(45)})  is
responsible  for  the  line  width  of  the  plasma-Bloch
resonance.

      We   have   calculated   the  hybrid   plasma-Bloch
oscillation frequency $f_H=\omega^H_+/2\pi$, 
using Eqs. (\ref{(45)}) and (\ref{(46)}), 
for the
typical  values of the superlattice 
parameters  
$\epsilon_0\simeq 13$, 
$d\simeq 50$ {\AA},
$E_P\simeq 10$ kV/cm,
$f_\nu = \nu/2\pi = 1.2$ THz for
different values of the current densities  $j_p$ (see Fig.\ \ref{Fig7}).  For
small  values of the current densities $j_P \simeq 10$ kA/cm$^2$
the frequency  of
the hybrid oscillation increases with applied voltage  in
all  range  of  the parameter
$V_{SL}/V_P$ . On the  other  hand,  for
higher  values  of  the  current  densities
$j_P\simeq (50 - 1000)$ kA/cm$^2$   the  hybrid
oscillation's   frequency   starts   to   decrease   with
increasing  bias  voltage  in the  sub-threshold  voltage
range  $V_{SL}\le V_P$.  Then, at super-threshold voltages
$V_{SL}\ge V_P$,  
$\omega_H$ starts  to
increase  again  tending to the Bloch  frequency.  It  is
important  to note that at high values of the dc  current
densities  $j_P$ the  hybrid plasma-Bloch oscillations  become
well  defined eigenmodes of the system ($f_H \ge f_\nu$). 
Therefore,  an
essential improvement of the matching efficiency  between
antenna and the superlattice can be expected in the high-
frequency  range  due to a resonant  excitation  of  this
eigenmode in the device.

\begin{figure}
\begin{center}
\epsfxsize=70mm
\epsffile{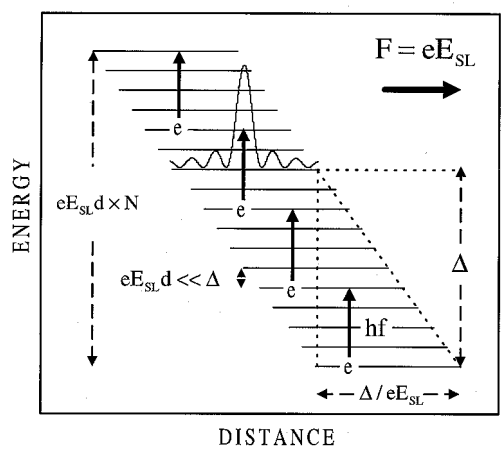}
\caption{
Real space energy diagram illustrating  THz-
photon  ($f\gg \nu/2\pi$)  detection  in the superlattice:  DC  electric
field  $E_{SL}$  is   applied   to  the  $N$-period   semiconductor
superlattice with the miniband width $\Delta$ . Under the  action
of the dc field electrons perform Bloch oscillations with
the  spatial amplitude $\Delta/eE_{SL}$ . At critical dc electric voltage
(field)
$V_{SL}=V_P=N\hbar\nu/e$ ($eE_{SL}d = \hbar\nu$)  
electrons move against the dc electric  force
due  to  absorption of photons climbing up  the  Wannier-
Stark  ladder.  The  energy $2eV_P$  should  be  absorbed   from
external ac field in order to subtract one electron  from
the  external circuit. One half of this energy is  needed
for  the electron to overcome the potential barrier which
is  formed  by  the dc force, with the other  half  being
delivered  to  the lattice due to energy  dissipation.  A
quasi-classical description of the process  is  valid  if
$f\ll \Delta/\hbar$
when  allowed transitions between different Wannier-Stark
state exist.}
\label{Fig8}
\end{center}
\end{figure}
\subsection{High-frequency limit}
     
Let  us  compare  the high-frequency  limit  of  the
responsivity  of  the  superlattice  with   the   quantum
efficiency $R_{\rm max} = e/\hbar\omega$    
which  is  believed  to  be  a  fundamental
restriction for the responsivity of superconductor tunnel
junctions\cite{ref18}.  This quantum efficiency (or quantum  limit)
corresponds to the tunnelling of one electron across  the
junction  for  each  signal  photon  absorbed\cite{ref18},  with  a
positive sign of the responsivity.

     In our case the mechanism of the photon detection is
different  (see  Fig.\  \ref{Fig8}).  Electrons  move  against  the
applied  dc electric force due to absorption of  photons.
At $V_{SL}=V_P$   the  responsivity is negative, indicating  that  one
electron is subtracted from the dc current flowing  through
the  superlattice when the energy  $2eV_P$ is absorbed  from  the
external ac field. One half of this energy is needed  for
the  electron to overcome the potential barrier which  is
formed by the dc force, with another half being delivered
to  the lattice due to energy dissipation. If the applied
dc  voltage is strong enough, i.e. $V_{SL}\gg V_P$, 
dissipation plays no
essential role in the superlattice responsivity. In  this
case the energy $eV_{SL}$ should be absorbed from the ac field  in
order to subtract one electron from the dc current simply
due to the energy conservation law.
\begin{figure}[!ht]
\begin{center}
\epsfxsize=70mm
\epsffile{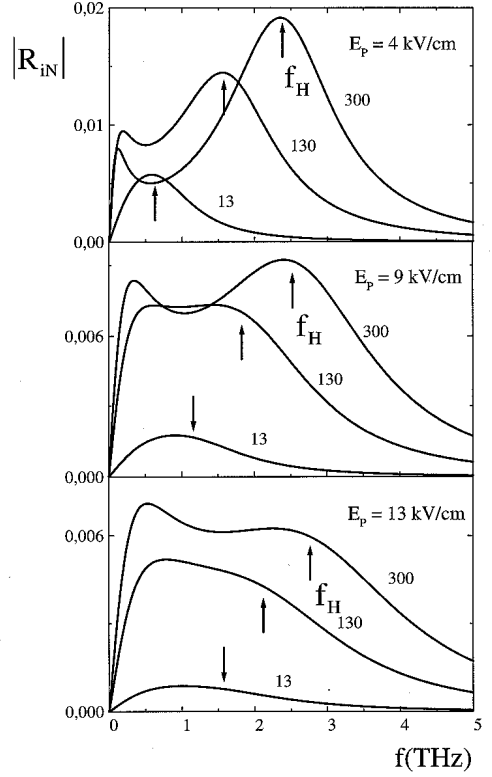}
\caption{The  frequency dependence of the  normalised
current  responsivity  
$|R_{iN}|=|R_i/(e/\hbar\omega)|$ 
of  the  superlattice  THz-photon detector 
($a=2\mu$m, 
$L=0.5\mu$m, 
$R_S=10\Omega$,
$V_{SL} = 0.95 V_P$)  for three values of  the
peak current density 
($j_P=$ 13, 30, and 300 kA / cm$^2$ ) and  for
three values of the peak electric field 
($E_P=$ 4, 9, and 13 kV/cm ) (from Ref.[24]).
The   relevant   positions  of  the  hybrid  plasma-Bloch
frequencies $f_H$  are  indicated for  each  curve  by  arrows
showing   characteristic  resonance  (high  peak  current
densities)  and  roll-off  (low peak  current  densities)
behavior.}
\label{Fig9}
\end{center}
\end{figure} 
\subsection{Excitation of the plasma-Bloch oscillations}

     For demonstration of the frequency dependence of the
superlattice  current responsivity in  the  THz-frequency
band we will focus on the GaAs/Ga$_{0.5}$Al$_{0.5}$As superlattices
specially   designed  to  operate  as   millimeter   wave
oscillators at room temperature. In Ref. \cite{ref43} 
wide-miniband
superlattice    samples   with
$d=50$ {\AA},
$\Delta\simeq 113$ meV,
$n\simeq 10^{17}$ ${\rm cm}^{-3}$,    were   investigated
experimentally. They demonstrated a well-pronounced Esaki-
Tsu  negative differential conductance for
$E_{SL}\ge E_P \simeq 4$ kV/cm  with the high
peak current of the order of 
$j_P \simeq 130$ kA/cm$^2$. The measured value of  the
peak  current  is in a good agreement with  the  estimate
$j_P\simeq(80-160)$ kA/cm$^2$
for  
$n\simeq (1-2)\times 10^{17}$,  
$T=$ 300 K  based  on Eq. (\ref{(12)}),  if  one  assumes  an
equilibrium   Boltzmann  distribution  for   the   charge
carriers.  From  the peak electric field and  current  we
find   the   scattering  and  plasma  frequencies  
$f_\nu \simeq 0.5$ THz, $f_P = 2$ THz,
respectively,   assuming   
$\epsilon_0 = 13$
for  the  average   dielectric
lattice   constant.   The  maximum  frequency   for   the
semiclassical approach to be valid for these samples is 
$f_\Delta \simeq 27$ THz.

     Figure  \ref{Fig9}  shows  the frequency  dependence  of  the
normalised  current  responsivity  calculated  for  three
values  of  the peak current density in the superlattice,
i.e. $j_P=$ 13, 130,  and  300
kA/cm$^2$ and for three values  of  the  peak
electric field,
$E_P=$ 4,  9, and
13 kV/cm.  We  also use the typical values for the superlattice
length $L=$ $0.5\mu$m   (superlattice  consists  of  100  periods),  and
assume $a=$ $2\mu$m  for the superlattice mesa 
radius \cite{ref43,ref44,ref45}. We choose $R_S$ $=10\Omega$
for  the  series  resistance of the device  in  the  THz-
frequency  band,  i.e.  the same value  as  for  resonant
tunnelling diodes having the same radius of mesas \cite{ref37}.  The
calculations are performed in the region of the  positive
differential conductance for dc bias voltage close to the
peak voltage ($V_{SL}=0.95$ $V_P$)

For $E_P = 4$ kV/cm  
($f_\nu\simeq 0.5$ THz) Fig.\ \ref{Fig9} demonstrates well-pronounced resonant
behavior of the normalised responsivity as a function of
frequency. The resonance frequency and the maximum  value
of  the  responsivity  rise if the peak  current  density
increases.  For 
$j_p = 300$ kA/cm$^2$ the normalised responsivity reaches  its
maximum value $-R_{iN}\simeq 0.02$
($-R_i \simeq 2$ A/W) at frequency $f\simeq 2.5$ THz.
    For higher values of the peak electric fields 
$E_P = 9$ kV/cm ($f_\nu\simeq 1.08$ THz) and
$E_P = 13$ kV/cm ($f_\nu\simeq 1.57$ THz)
the  resonance  line-widths  are  broadened  due  to
implicit  increase  of  the  scattering  frequencies.  In
particular,  for
$E_P = 13$ kV/cm,
$j_P=300$ kA/cm$^2$  the normalised responsivity  has  an
almost  constant  value  
$-R_{iN}\simeq 0.006$
($-R_i \simeq 0.6$ A/W) up  to $f\simeq 2.5$ THz   and,  then,  rapidly
decreases.
      The   frequency   behavior   of   the   normalised
responsivity  originates from excitation of  the  plasma-
Bloch  oscillations in the superlattice. We  indicate  in
Fig.  \ref{Fig9}  the  positions of the hybrid  frequencies
$f_H=|\omega^H_\pm|/2\pi$   with
arrows. For small peak electric fields (low values of the
scattering  frequencies) the hybrid frequency corresponds
to the maximum of the normalised responsivity. For higher
values of the peak field (higher values of the scattering
frequencies) it corresponds to the roll-off frequency  at
which the responsivity starts to decline.

\subsection{Optimized superlattice length}
      The  enhancement  of  the  normalised  responsivity
requires   an   optimum  matching   efficiency   of   the
superlattice  to the broad-band antenna and  minimization
of  the  parasitic losses in the series  resistor.  These
requirements impose an optimum length of the superlattice
for each chosen frequency of the incoming THz-photons and
series resistance.
\begin{figure}[thbp]
\begin{center}
\epsfxsize=70mm
\epsffile{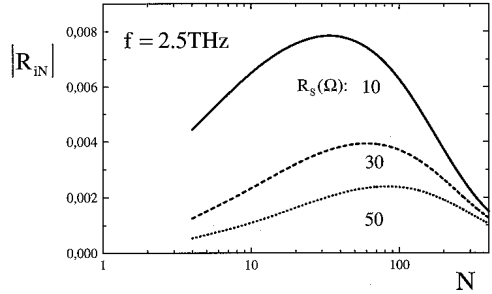}
\caption{The  dependence  of the  normalised  current
responsivity  
$|R_{iN}|=|R_i/(e/\hbar\omega)|$
of the superlattice THz-photon detector 
($a=2\mu$m,
$E_P=4$kV/cm,
$j_P=130$ kA/cm$^2$) at 
$f =2.5$ THz for three values of the
series  resistance  
($R_S$=10, 30, and 50 $\Omega$)  as  a  function  of
number of the superlattice periods $N$ (from Ref.[24]).}
\label{Fig10}
\end{center}
\end{figure}

     We  show  in Fig.\ \ref{Fig10} the dependence of the normalised
responsivity  on  the number of the superlattice  periods
for $f= 2.5$ THz.We used for calculation 
$a=2\mu$m,
$j_P = 130$ kA/cm$^2$,
$V_{SL} = 0.95$  $V_P$,  and three values  of
the  series resistance $R_S = 10$, 30 and 50 $\Omega$. 
For all three  values
of the series resistance the responsivity displays a well
pronounced  maximum  for  the  optimum  number   of   the
superlattice  periods  $N=N_{\rm max}$. The value  of
$N_{\rm max}$   increases  with
increasing  of the series resistance 
($N_{\rm max}\simeq 40$ for $R_S$ = 10 $\Omega$,
$N_{\rm max}\simeq 60$ for $R_S$ = 30 $\Omega$  and
for 
$N_{\rm max}\simeq 90$ for $R_S$ = 50 $\Omega$). 
This result can be readily understood by recalling
that  a  larger  volume  of  the  superlattice  minimizes
parasitic   losses  for  higher  values  of  the   series
resistance   because  of  reduction   of   the   sample's
capacitance.  

In addition to the examples discussed here,
Ref. \cite{JAP} reports several other aspects of the
superlattice responsivity, such as bias voltage dependence,
optimized peak current density etc.

\subsection{Conclusions}
We have illustrated here the steps required to perform a
superlattice device optimization.  These include: (i) Consideration
of the frequency dependence of the superlattice.  Here a truly
microscopic analysis is not yet available, but a Boltzmann equation
based theory should be applicable, with due caution. (ii) The impedance
matching between the antenna and the superlattice must be optimized.
(iii) The effect of parasitic losses must be included in the
analysis.  One of the main results emerging from our analysis is
the importance of collective excitations: the hybridized plasma-Bloch
oscillations both enhance the responsivity and increase the roll-off
frequency.  It is found that an optimized superlattice can have a
responsivity which approaches 10 \% of an ideal superconducting tunnel
junction.  The great advantage, of course, is that superlattice based
detectors work even at room temperature, and this property should offer
a wide range of applications.


\begin{thebibliography}{}

\bibitem{1.}{ESA70}{[1]}
L. Esaki and R. Tsu, IBM~J.~Res.~Develop. {\bf 14},  61  (1970).

\bibitem{2.}{SIB90}{[2]}
A. Sibille, J.~F. Palmier, H. Wang, and F. Mollot, Phys.~Rev.~Lett. {\bf 64},
  52  (1990).

\bibitem{3.}{GRA91}{[3]}
H.~T. Grahn, R.~J. Haug, W. M{\"u}ller, and K. Ploog, Phys.~Rev.~Lett. 
{\bf  67},  1618  (1991).

\bibitem{4.}{WAS93}{[4]}
C. Waschke {\it et~al.}, Phys.~Rev.~Lett. {\bf 70},  3319  (1993).

\bibitem{5.}{HOL92}{[5]}
M. Holthaus, Phys.~Rev.~Lett. {\bf 69},  351  (1992).

\bibitem{6.}{KEA95b}{[6]}
B.~J. Keay {\it et~al.}, Phys.~Rev.~Lett. {\bf 75},  4102  (1995).

\bibitem{7.}{LEB70}{[7]}
P.~A. Lebwohl and R. Tsu, J.~Appl.~Phys. {\bf 41},  2664  (1970).

\bibitem{8.}{TSU75}{[8]}
R. Tsu and G. D{\"o}hler, Phys.~Rev.~B {\bf 12},  680  (1975).

\bibitem{9.}{MIL94}{[9]}
D. Miller and B. Laikhtman, Phys.~Rev.~B {\bf 50},  18426  (1994).

\bibitem{10.}{WAC97b}{[10]}
A. Wacker and A.~P. Jauho, Physica~Scripta {\bf T69},  321  (1997).

\bibitem{11.}{Aguado}{[11]}
R. Aguado, G. Platero, M. Moscoso, and L.~L. Bonilla,
Phys. Rev. B {\bf 55}, 16053 (1997)
\bibitem{12.}{prl1}{[12]}
A. Wacker and A.~P. Jauho, Phys.~Rev.~Lett. {\bf 80}, 369 (1998).

\bibitem{13.}{prl2}{[13]}
A. Wacker, A.~P. Jauho, S. Rott. A. Markus, P. Binder, and G.~H. D{\"o}hler,
Phys.~Rev.~Lett. {\bf 83}, 836 (1999).

\bibitem{14.}{IGN91}{[14]}
A.~A. Ignatov, E.~P. Dodin, and V.~I. Shashkin, Mod.~Phys.~Lett.~B {\bf 5},
  1087  (1991).

\bibitem{15.}{LEI91}{[15]}
X.~L. Lei, N.~J.~M. Horing, and H.~L. Cui, Phys.~Rev.~Lett. {\bf 66},  3277
  (1991).

\bibitem{16.}{ROT97}{[16]}
S. Rott, N. Linder, and G.~H. D{\"o}hler, Superlattices and Microstructures
  {\bf 21},  569  (1997).

\bibitem{17.}{BRY97}{[17]}
V.~V. Bryksin and P. Kleinert, J.~Phys.: Cond.~Mat. {\bf 9},  7403  (1997);
S. Rott {\it et al.}, Physica E (Amsterdam) {\bf 2}, 511 (1998).

\bibitem{18.}{MAH90}{[18]}
G.~D. Mahan, {\em Many-Particle Physics} (Plenum, New York, 1990).

\bibitem{19.}{MUR95}{[19]}
S.~Q. Murphy, J.~P. Eisenstein, L.~N. Pfeiffer, and K.~W. West, Phys.~Rev.~B
  {\bf 52},  14825  (1995).

\bibitem{20.}{WAC97}{[20]}
A. Wacker,  in {\em Theory of transport properties of semiconductor
  nanostructures}, edited by E. Sch{\"o}ll (Chapman and Hall, London, 1998),
  Chap. 10.
  
\bibitem{21.}{HAU96}{[21]}
H. Haug and A.-P. Jauho, {\em Quantum Kinetics in Transport and Optics of
  Semiconductors} (Springer, Berlin, 1996).

\bibitem{22.}{ref35}{[22]}  
A.~A. Ignatov, E. Schomburg, J. Grenzer, S. Winnerl,
K.~F.   Renk   and   E.~P.  Dodin,   Superlattices   \&
Microstructures {\bf 22}, 15 (1997).

\bibitem{23.}{ref18}{[23]} 
For a review, see J. R. Tucker and M. J. Feldman, Rev.
Mod. Phys. {\bf 57}, 1055 (1985).

\bibitem{24.}{JAP}{[24]}
A.~A. Ignatov and A.~P. Jauho,
J. Appl. Phys. {\bf 85}, 3643 (1999)

\bibitem{25.}{ref37}{[25]}  
T.~C.~L.~G. Sollner, W.~D. Goodhue, P.~E. Tannenwald,
C.~D. Parker, and D.~D. Peck, Appl. Phys. Lett. {\bf 43},  588
(1983).

\bibitem{26.}{ref38}{[26]}
 H.~C.  Torrey and C.~A. Whitmer, Crystal  Rectifiers
(McGraw-Hill, New York, 1948), p. 336.

\bibitem{27.}{ref39}{[27]} 
A.~A. Ignatov and V.~I. Shashkin,
Sov. Phys. JETP. {\bf 66},
526 (1987).

\bibitem{28.}{ref41}{[28]}  
R.~G. Chambers, Proc. Phys. Soc. (London) A {\bf 65},  458
(1952).

\bibitem{29.}{ref2}{[29]} 
R. Tsu and L. Esaki, Appl. Phys. Lett. {\bf 19}, 246 (1971).

\bibitem{30.}{ref3}{[30]}  
A.~A.  Ignatov and Yu.~A. Romanov, Sov.  Phys.  Solid
State  {\bf 17},  2216 (1975); 
Phys. Status Solidi  B  {\bf 73},  327
(1976).

\bibitem{31.}{ref5}{[31]} 
 A.  A.  Ignatov  and Yu. A. Romanov, Radiophysics  and
Quantum Electronics (Consultants Bureau, N.Y., 1978) 
Vol. {\bf 21}, p. 90.

\bibitem{32.}{ref7}{[32]} 
M. Holthaus, Phys. Rev. Lett. {\bf 69}, 351 (1992).

\bibitem{33.}{ref17}{[33]} 
A. A. Ignatov, K. F. Renk, and E. P. Dodin, Phys. Rev.
Lett. {\bf 70}, 1996 (1993);
J. B. Xia, Phys. Rev. B {\bf 58}, 3565 (1998).

\bibitem{34.}{ref13}{[34]}  
A.  A. Ignatov, E. Schomburg, J. Grenzer, K. F. Renk,
and E. P. Dodin, Z. Phys. B {\bf 98}, 187 (1995).

\bibitem{35.}{ref43}{[35]} 
E. Dutisseuil, A. Sibille, J.~F. Palmier, F. Aristone,
F.  Mollot,  and V. Thietty-Mieg, Phys. Rev. B  {\bf 49},  5093
(1994).

\bibitem{36.}{ref44}{[36]}  E.  Schomburg, A.~A. Ignatov, J. Grenser,
K.~F. Renk,
D.~G.  Pavel'ev,  Yu. Koschurinov,  B.~Ja.  Melzer,  S.
Ivanov, S. Schaposchnikov, and P.~S. Kop'ev, Appl.  Phys.
Lett. {\bf 68}, 1096 (1996).

\bibitem{37.}{ref45}{[37]}  S. Winnerl, E. Schomburg, J. Grenser, H.-J. Regl, 
A.~A. Ignatov, A.~D. Semenov,
K.~F.  Renk,  D~ G. Pavel'ev, Yu. Koschurinov,  B.~Ja.
Melzer, V. Ustinov, S. Ivanov, S. Schaposchnikov, and  
P.~S. Kop'ev, Phys. Rev. B {\bf 56}, 10 303 (1997).


\end{thebibliography}
\end{document}